\documentclass[fleqn,usenatbib]{mnras}

\usepackage{newtxtext,newtxmath}

\usepackage[T1]{fontenc}

\DeclareRobustCommand{\VAN}[3]{#2}
\let\VANthebibliography\thebibliography
\def\thebibliography{\DeclareRobustCommand{\VAN}[3]{##3}\VANthebibliography}

\usepackage{graphicx}	
\usepackage{amsmath}	
\usepackage{amssymb}	
\usepackage{multirow}
\usepackage{enumitem}
\usepackage{booktabs}
\usepackage{graphicx}
\usepackage{bm}

\def\lcdm{$\Lambda$CDM}
\def\hc{$H_0$}
\def\tdd{$D_{\Delta t}$}
\def\hst{{\it HST}}
\def\efr{{$R_{\mathrm{eff}}$}}

\def\kext{\ensuremath{\kappa_{\rm ext}}}
\def\jcap{Journal of Cosmology and Astroparticle Physics}
\newcommand{\hcuint}{km~s$^{-1}$~Mpc$^{-1}$}
 
\title[TDLMC. I]{Time Delay Lens Modeling Challenge: I. Experimental Design}

\author[X. Ding et al.]{
Xuheng Ding,$^{1,2}$\thanks{dxh@astro.ucla.edu}
Tommaso Treu,$^{1}$
Anowar J. Shajib,$^{1}$
Dandan Xu,$^{3}$
Geoff C.-F. Chen,$^{4}$ \newauthor
Anupreeta More,$^{5}$
Giulia Despali,$^{6}$
Matteo Frigo,$^{6}$
Christopher D. Fassnacht,$^{4}$\newauthor
Daniel Gilman,$^{1}$
Stefan Hilbert,$^{7,8}$
Philip J. Marshall,$^{9}$
Dominique Sluse,$^{10}$\newauthor
Simona Vegetti$^{6}$
\\
$^{1}${Department of Physics and Astronomy, University of California, Los Angeles, CA, 90095-1547, USA}\\
$^{2}${School of Physics and Technology, Wuhan University, Wuhan 430072, China}\\
$^{3}${Department of Astronomy, Tsinghua University, Beijing, 100084, China}\\
$^{4}${Department of Physics, University of California, Davis, CA 95616, USA}\\
$^{5}${Kavli IPMU (WPI), UTIAS, The University of Tokyo, Kashiwa, Chiba 277-8583, Japan}\\
$^{6}${Max Planck Institute for Astrophysics, Karl-Schwarzschild-Strasse 1, D-85740 Garching, Germany}\\
$^{7}${Exzellenzcluster Universe, Boltzmannstr. 2, 85748 Garching, Germany}\\
$^{8}${Ludwig-Maximilians-Universit{\"a}t, Universit{\"a}ts-Sternwarte, Scheinerstr. 1, 81679 M{\"u}nchen, Germany}\\
$^{9}${Kavli Institute for Particle Astrophysics and Cosmology, Stanford University, 452 Lomita Mall, Stanford, CA 94035, USA}\\
$^{10}${STAR Institute, Quartier Agora - All\'ee du six Ao\^ut, 19c B-4000 Li\`ege, Belgium}
}

\date{Accepted XXX. Received YYY; in original form ZZZ}

\pubyear{2020}

\begin{document}
\label{firstpage}
\pagerange{\pageref{firstpage}--\pageref{lastpage}}
\maketitle

\begin{abstract}
Strong gravitational lenses with measured time delay are a powerful
tool to measure cosmological parameters, especially the Hubble
constant (\hc). Recent studies show that by combining just three
multiply-imaged AGN systems, one can determine \hc\ to 2.4\%
precision. Furthermore, the number of time-delay lens systems is
growing rapidly, enabling, in principle, the determination of \hc\
to $1\%$ precision in the near future. However, as the precision
increases it is important to ensure 
that systematic errors and biases remain
subdominant. For this purpose, challenges with simulated datasets are a key component in
this process. Following the experience of the past challenge on time delay,
where it was shown that time delays can indeed be measured
precisely and accurately at the sub-percent level, we now present the
``Time Delay Lens Modeling Challenge'' (TDLMC). The goal of this challenge is
to assess the present capabilities of lens modelling codes and
assumptions and test the level of accuracy of inferred cosmological
parameters given realistic mock datasets. We invite scientists to model a
set of simulated  \textit{Hubble Space Telescope} (\hst) observations of 50 mock lens
systems. The systems are organized in rungs, with the complexity and realism
increasing going up the ladder.
The goal of the challenge is to infer \hc\ for each rung,
given the \hst\ images, the time delay, and a stellar velocity
dispersion of the deflector, for a fixed background cosmology.
The TDLMC challenge starts with the mock data release on 2018
January 8th.
The deadline for blind submission is different for each rung.
The deadline for Rung~0-1 is 2018 September 8;
the deadline for Rung~2 is 2019 April 8 and the one for Rung~3 is 2019 September 8.  
This first paper gives an overview
of the challenge including the data design, and a set of metrics to
quantify the modelling performance and challenge details. After the
deadline, the results of the challenge will be presented in a
companion paper with all challenge participants as co-authors.
\end{abstract}




\section{Introduction}
\label{sec:introduction}
During the past decade, the flat $\Lambda$Cold Dark Matter (\lcdm) model has provided an accurate description of the geometry and dynamics of our Universe. This model, now referred to as the standard model, has demonstrated an excellent fit to variety of independent of cosmological observations including the analysis of cosmic microwave background (CMB) by the \textit{Planck} and \textit{WMAP} satellies \citep{Pla++13, Planck2015XIII} and low redshift cosmic probes such as type Ia supernovae \citep{Rie++16, Betoule2014}, baryon acoustic oscillation (BAO) surveys \citep{Eisenstein2005, Alam2017}, cosmic shear \citep{Kilbinger2015}, and the gas fraction in clusters of galaxies \citep{Mant++10,Mantz2010a}. Interestingly, however, in the flat \lcdm\ model the Hubble Constant (\hc) inferred from the extrapolation of the \textit{Planck} measurements at high redshift is in tension with the local measurement from the traditional cosmic distance ladder \citep{Rie++16}. If this tension were confirmed at higher significance, it would be a major discovery, requiring deviations from the standard flat \lcdm\ and possibly new physics.
Thus, improving the precision of the measurement of \hc\ is a central goal of current cosmological efforts. On the one hand, it is important to improve the quality of each method. On the other hand, it is essential to develop independent methods, providing an independent check on potential systematic errors.

In the past few years, it has been shown that strongly
lensed AGN with measured time delays can constrain \hc\ with $\sim$5\% precision
per system, given high-quality data and state-of-the-art
modelling techniques \citep{Suy++10, Suy++13,Suy++14}. 
Recently, the \hc\ Lenses in COSMOGRAIL's Wellspring (H0LiCOW) and SHARP collaborations have combined the analysis of six strong lensed quasars and obtain a joint inference of \hc\ to 2.4\% level as $H_0 = 73.3\substack{+1.7\\-1.8}$ \hcuint ~\citep{Wong2019, h0licow5, Birrer2019, Chen2019, Rusu2019}. Additionally, the STRong-lensing Insights into the Dark Energy Survey (STRIDES) collaboration infers $H_0 = 74.2_{-3.0}^{+2.7}$ \hcuint with $3.9\%$ uncertainty by analyzing a new lensed quasar~\citep{Shajib2020}.
Going forward, the Time-Delay COSMOgraphy (TDCOSMO) collaboration, which combines the H0LiCOW/SHARP/COSMOGRAIL/STRIDES, aims to to measure \hc\ to
sub-percent level \citep{Millon2019, T+M16}.

Whereas analyzing increasingly large samples of strongly lensed AGNs is sufficient to meet the precision goal, it is also  crucial to make sure that the measurement is accurate, i.e. it does not suffer from systematic errors that may ultimately provide a noise floor or a bias. For time delay cosmography, systematic errors include both the known unknowns \citep[e.g., time delay measurement, residual uncertainties of the lens model and the line-of-sight structure, see][]{T+K2018, XuEtal2016, S+S13} and unknown unknowns. While an effective strategy to uncover the latter is performing blind analyses on the real data and checking the mutual consistency \citep{Suy++13,h0licow5}, the former can be quantified and hopefully corrected for by means of a series of dedicated challenges. 

In the past, the accuracy on the measurement of time delay has been estimated via a ``Time Delay Challenge'' (TDC) in which the realistic mock `observed' lensed AGN light curves were generated and then analyzed by the invited modelling teams \citep{TDC1, TDC2}. The mock light curves were modelled through a \textit{blind analysis}, where the true value of the mock time delay were unknown to the participating team. This strategy is crucial to avoid (unconscious) experimenter bias or reverse engineering efforts. In the end, \cite{TDC2} concluded that with light curves of sufficient quality, achievable with present-day technology time delays can indeed be measured with sub-percent accuracy and precision. \cite{TDC2} also estimated that under the most favorable assumptions the Large Synoptic Survey Telescope (LSST) should provide around 400 robust time-delay measurements with precision within 3\% and accuracy within 1\%. More work also needs to be done to address the effects of microlensing on time delay measurements \citep{T+K2018}. A new challenge, including multi-band data and fine microlensing-induced perturbation, is currently derived to provide a new benchmark for time-delay measurements (Liao et al. in prep).

A second potentially limiting source of systematic errors is the
inference of the lensing potential of the main deflector.  Even though
blind measurements of current samples demonstrate that the lens mass
models are sufficiently well constrained at a level of a few percent
given high signal-to-noise imaging and stellar velocity dispersion
\citep{Suy++13,Suy++14, h0licow4}, it is yet to be demonstrated that
the current approaches are sufficient to reach 1\% precision and accuracy.

Demonstrating this goal requires a dedicated effort, specific to the issue of Fermat potential. This is the topic  
%
%
%
of this paper and its companion presenting the ``Time
Delay Lens Modeling Challenge'' (TDLMC).

In this challenge, we
(hereafter ``Evil'' Team) provide realistic simulated time-delay lens
data including i) \hst-like lensed AGN images, ii) lens time delays,
iii) line-of-sight velocity dispersions and iv) external convergence to
the participating modelling teams (hereafter ``Good'' Teams)\footnote{We
  stress here that we follow the tradition of TDC and use the
  nicknames ``Evil'' and ``Good'' Teams. These nicknames do not
  denote any despicable intention or moral judgment, but were chosen
  to capture the desire of the challenge designers to produce
  significantly realistic (and difficult) lens data as well as an
  incentive for the outside teams to participate.}.  Likewise, blind
analysis is employed to assess the accuracy of lens modelling and
cosmological inference.  We emphasize that TDLMC is purely a lens
modelling challenge. In order to isolate the components of the time
delay cosmography measurement, we assume here that the time delays are
known precisely and accurately, and we only consider a single plane
deflector. Separate challenges have dealt and will deal with the
other elements. For simplicity of analysis, we also keep all the
cosmological parameters fixed except for \hc.

This paper is structured as follows. Section~\ref{sec:theory} briefly reviews the lens theory and introduces the ingredients used for the simulations. Section~\ref{sec:structures}, describes the simulated data sets and layout of this challenge. Section~\ref{sec:participation}, introduces four metrics aimed at evaluating the performance of the modelling result, and gives instructions to access the mock data and the timeline for the challenge.  Section~\ref{sec:sum} concludes with a short summary.

\section{Lens theory and Ingredients of the Simulations}
\label{sec:theory}
We briefly review the relevant strong lensing theory in Section~\ref{sec:theo}, and introduce the key ingredients including deflector/source surface brightness and deflector mass for simulating the lens image in Sections~\ref{sec:sruf_b} and \ref{sec:defl_m}, respectively.  
	
\subsection{Strong gravitational lensing}
\label{sec:theo}

For a strong lens system, the scaled deflection angle
of a light ray is $\bm{\alpha} = \vec{\nabla} \psi(\bm{\theta})$ and the deflection of light rays
can be described by the lens equation $\bm{\beta} = \bm{\theta} - \bm{\alpha}(\bm{\theta})$,
where $\psi(\bm{\theta})$ is the lens potential at position $\bm{\theta}$ on the plane of the sky (image plane) and
$\bm{\beta}$ is the source position in the absence of a deflector (source plane).

The traveling time from the source to the observer depends on both the path of the source light and on the lens gravitational potential of the deflector. These two effects lead to a difference in arrival time for the multiple images.
In theory, the time delay $\Delta t_{ij}$ between two lensed images is given as follows:
\begin{eqnarray}\label{eq:td}
&\Delta t_{ij} = \frac{D_{\Delta t}}{c} \left[
\phi(\bm{\theta}_i)-
\phi(\bm{\theta}_j)
\right],\\
\label{eq:fermat}
& \phi (\bm{\theta}_i)=\frac{(\bm{\theta}_i - \bm{\beta})^2}{2} -
\psi(\bm{\theta}_i),
\end{eqnarray}
where $\bm{\theta}_i$ and $\bm{\theta}_j$ are the coordinates of the images $i$ and $j$ in the image plane.
$\phi (\bm{\theta}_i)$ is the so-called Fermat potential and
$D_{\Delta t}$ is so-called time-delay distance, defined as:
\begin{equation}\label{eq:tdd}
D_{\Delta t} \equiv (1+z_d)
\frac{D_{\rm d} D_{\rm s}}{ D_{\rm ds}}.
\end{equation}
Here, $D_{\rm d}$, $D_{\rm s}$ and $D_{\rm ds}$ are respectively the angular
distances from the observer to the deflector, from the observer to the source, and from the deflector to the source. Thus, the time-delay distance is proportional to the inverse of the Hubble constant, i.e. $H_0^{-1}$.
By modelling the lens image of the time-delay lens, one can derive the Fermat potential,
deduce the \tdd\ and thus infer the value of the \hc\ (and other cosmological parameters).

The projected dimensionless surface mass density $\kappa(\bm{\theta})$ is:
\begin{equation}\label{eq:kappa}
\kappa(\bm{\theta})=\frac{1}{2}\nabla^2\psi(\bm{\theta}),
\end{equation}
and 
\begin{equation} \label{eq:f_kappa} 
\kappa(\bm{\theta}) = \frac {\Sigma(D_{\rm d} \bm{\theta})} {\Sigma_{\rm cr}} \qquad \mathrm{with} \qquad \Sigma_{\rm cr} = \frac{c^2 D_{\rm s}}{4 \pi G D_{\rm d} D_{\rm ds}},
\end{equation}
where $\Sigma(D_{\rm d} \bm{\theta})$ is the physical projected surface mass density of the deflector and $\Sigma_{\rm cr}$ is the critical surface density.


\subsection{Surface brightness}
\label{sec:sruf_b}

To study how results change with increasing complexity, we adopted a variety of approaches to simulate the brightness profiles of the lens and source galaxies.

\subsubsection{S\'ersic model}

As a matter of convenience, in the entry level of the challenge, we choose a
common simply-parameterized description of the surface brightness for the
lens and source galaxy. This choice is meant primarily for testing the
codes, both for ``Evil'' and ``Good'' Teams.  In the literature, the
S\'ersic profile \citep{Ser68} is one of the most commonly used models
to describe the surface brightness of galaxies. It ranges from
exponential discs to \citet{deV48} profiles.

The S\'ersic profile is parameterized by:
\begin{eqnarray}
   \label{eq:sersic}
   &I(R) = A \exp\left[-k\left(\left(\frac{R}{R_{\mathrm{eff}}}\right)^{1/n}-1\right)\right] ,\\
   &R(x,y,q) = \sqrt{qx^2+y^2/q}.
\end{eqnarray}
where $A$ is the amplitude and S\'ersic index $n$ controls the shape of the radial
surface brightness profile; a larger $n$ corresponds to a steeper
inner profile and a highly extended outer wing. 
 $k$ is a constant which
depends on $n$ so as to ensure that the isophote at $R=$\efr\ 
encloses half of the total light \citep{C+B99} and
$q$ denotes the axis ratio.

\subsubsection{Realistic source image as AGN host}

\begin{figure}
\centering
{\includegraphics[trim = 0mm 20mm 0mm 0mm, clip,width=0.35\textwidth]{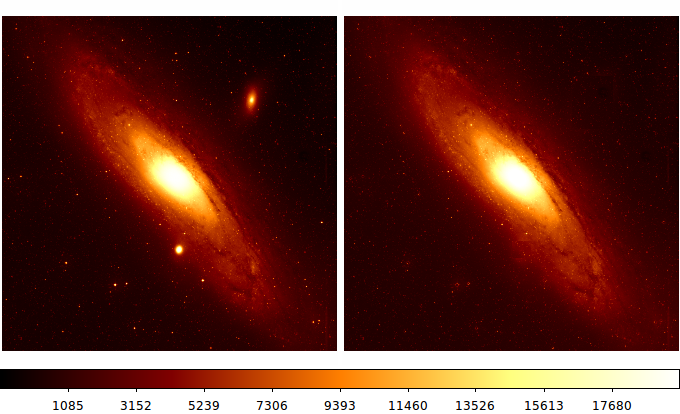}}
\caption{\label{fig:real_source}
Illustration of real \hst\ galaxy image as lensed source host. The bright point source and foreground/background galaxies in the original image (left) are replaced by interpolation of nearby pixels to obtain a clean galaxy image (right) .}
\end{figure}

For the bulk of the challenge, we use more realistic and complex surface brightness distribution for the host galaxy of the lensed AGN. For example, we use real images of galaxies appropriately smoothed and cleaned by foreground/background contaminants as shown in  Fig.\ref{fig:real_source}.

\subsection{Deflector mass}
\label{sec:defl_m}

Likewise, we achieve different levels for complexity of the challenge by increasing the realism of the deflector mass distribution stepping up the ladder.  

\subsubsection{Elliptical power-law mass distribution}
A common simply-parameterized description of the deflector mass density profile is given by elliptical power-law models whose surface mass density is given by:
\begin{equation}
 \label{massmodel}
 \Sigma(x,y)=\Sigma_{cr}\frac{3-\gamma^{\prime}}{2}\left(\frac{\sqrt{q_m x^{2}+y^{2}/q_m}}{R_{\rm E}}\right)^{1-\gamma^{\prime}},
\end{equation}
$q_m$ described the projected axis ratio.
The so-called Einstein radius $R_{\rm{E}}$ is chosen such
that, when $q_m=1$ (i.e. spherical limit), it encloses a mean surface
density equal to $\Sigma_{cr}.$
The exponent $\gamma'$ is the slope of the power-law profile,
for massive elliptical galaxies $\gamma' \approx2$  \citep{T+K02a,T+K04,Koo++09}.
 We refer the reader to the reviews by \citet{Sch06, Bar10, Tre10} for more details.

\subsubsection{Simulated realistic galaxy mass distribution}
\label{sec:Sim_mass}

In order to achieve a more realistic deflector mass distribution, we also consider massive early-type lens galaxies produced by cosmological numerical simulations.  We only consider a single deflector, and do not include the effects of the line of sight other than via the external convergence introduced before. We choose systems with virial mass approximately 10$^{13}$ M$_\odot$ yielding Einstein radii of order $1''$ for typical source and deflector redshift ($z_d\approx0.5$) and ($z_s \approx1.5$).




\section{Structure of the Challenge}
\label{sec:structures}

In this section, we first describe the data sets that are made available to the ``Good'' Teams in Section~\ref{sec:sim_data}. Then, a description of the layers (rungs) of the challenge is given in Section~\ref{sec:rung}.

\subsection{Data sets}	
\label{sec:sim_data}
The mock data available to the ``Good'' teams consist of  deep \hst\ images, time delays, stellar velocity dispersion, and external convergence, as described below.
The released mock data sets have been tested by analyzing a subset of them with two independent lens modelling software and verifying that the input cosmology (and lens parameters when applicable) could be recovered within the uncertainties.


\subsubsection{\hst\ Image}
In order to mimic a typical observational setup in state of the art
observations, we choose to simulate high-resolution images obtained
with the \textit{Hubble Space Telescope} (\hst), using the Wide Field
Camera 3 (WFC3) IR channel in the F160W band. Even though this setup
has lower resolution than optical images taken with WFC3-UVIS or ACS,
we adopt it in order to minimize the effects of dust extinction and
optimize the contrast between the (blue) AGN and (red) host
galaxy. We adopt a range of AGN to host flux ratios so as to produce a distribution similar to that observed in real systems \citep{H0licow6,H0licow7}.
For simplicity, we do not include any dust extinction, and we assume AGN to be at the center of the host galaxy.
Also, multi-band datasets or adaptive
optics assisted ground-based images are left for future challenges.

In practice, the following steps are taken in order to simulate
realistic lens configurations.

\begin{figure}
\centering
{\includegraphics[trim = 0mm 20mm 0mm 0mm, clip,width=0.45\textwidth]{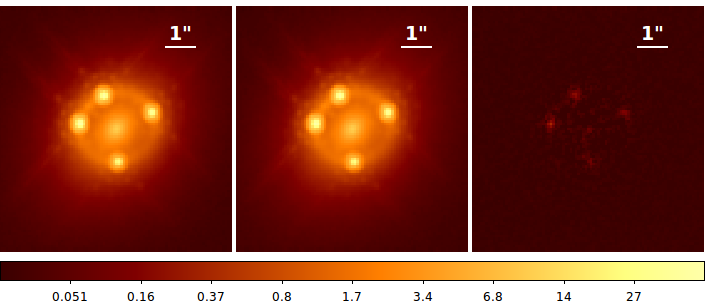}}
\caption{\label{fig:sim-images}
The left and middle panels illustrate the simulated \hst-like images based on two independent codes with same lens parameters. The right panel shows the difference on the same scale. The pixel scale is 0\farcs08 after drizzling.}
\end{figure}

\begin{enumerate}
\item For every set of lens and source parameters, compute high-resolution images of the lensed host, and deflector light. 
\item Convolve with the point spread function (PSF) appropriate for WFC3/F160W.
\item Compute the image plane positions and fluxes of the lensed AGN images and add them as appropriately scaled PSF in the image plane.
\item Rebin the oversampled images to the actual data resolution. Using different rebinning patterns, one can simulate eight dithering images in order to drizzle them\footnote{{\sc MultiDrizzle} is adopted for the drizzling, see \url{http://www.stsci.edu/hst/wfpc2/analysis/drizzle.html} for more information.} in step (vi).
\item Add noise based on realistic observation condition, including
  background, readnoise and Poisson noise from the source. The
  exposure time of each one of the eight images is taken to be 1200s, thus the
  total exposure time is 9600s.
\item Drizzle the individual images to recover
some of the resolution lost due to pixelization. Following common practice, we drizzle eight images into one final image; the corresponding pixel size is 0\farcs13 and 0\farcs08, before and after drizzling. 
 This step introduces correlated noise.  In order to allow ``Good'' Teams to model the original 
data, the eight non-drizzled images of one lens system are provided 
in addition to the final drizzled image.
\end{enumerate}
A detailed description of these steps is given by
\citet[][Section~3]{H0licow6}. In order to control for numerical
issues and for implicit bias in favor of any ``Good'' Team, we use two
independent codes to generate the simulations (half the sample with
each code).  An example of mock images generated by two independent
codes with the same parameters is shown in Fig.~\ref{fig:sim-images}.
The noise maps for the images are provided which contain the standard deviation of the noise.

\subsubsection{Time Delay}

Once the values of lens parameters are set, the difference of the Fermat potential between the AGN images can be calculated with Eq.~(\ref{eq:fermat}). To calculate the corresponding time delay with Eq.~(\ref{eq:td}), we need to assume a set of cosmological parameters. For simplicity, we draw values randomly from a uniform distribution between 50 to 90~km~s$^{-1}$~Mpc$^{-1}$ for the Hubble Constant, assuming a flat $\Lambda$CDM cosmological model with ${\Omega}_{\rm{m}}=1-{\Omega}_{\Lambda}=0.27.$ 

We then add measurement uncertainty to the time delay. As discussed in the introduction, the time delay is supposed to be known with sufficient precision and accuracy so that we can test the precision and accuracy of the models. We thus assume zero bias and the smallest random errors that can be obtained with current monitoring strategies. Thus we adopt as random error the largest between 1\% and 0.25 days.

\subsubsection{External Convergence}
\label{kext}
In principle, all mass along the line-of-sight contributes to the
deflection of light rays. In practice, however, it is often the case
that the lensing configuration can be approximated by a single main
deflector with the addition of external shear and convergence
(\kext). The latter is particularly important because it does not
change the image positions and relative fluxes, but it affects the
relative Fermat potential, and hence the time delay, according to the
following equation.
\begin{equation}\label{eq:k_ext}
\Delta t_{\rm obs} = (1-\kappa_{\rm ext}) \Delta t_{\rm true} .
\end{equation}
If not accounted for, the \kext\ can bias the inference of time delay
distance and thus \hc. A common practice to constrain the \kext\ is to
compare the distribution of mass along the line-of-sight with
numerical simulations \citep{h0licow3, Gre++13,Col++13, Suy++10,Collett2013, Hil++09}.

Since the focus of this challenge is single plane lens modelling, we
include the effects of the line-of-sight contribution in the following
simplified manner. We randomly generate a \kext\ using a random
Gaussian distribution with 0 and 0.025 as mean value and standard
deviation, respectively. This, from the point of view of modelling, one
can adopt a prior on \kext\ of $0\pm0.025$. The uncertainty is chosen
to represent well-characterized lines of sight, and corresponds to a
random uncertainty of 2.5\% on \hc, to first approximation.

\subsubsection{Lens Velocity Dispersion}
\label{sec:vd}
Stellar kinematic information is essential for breaking the mass-sheet
degeneracy and constraining the lensing potential.  Furthermore, the
measurement of stellar kinematics provides extra cosmological information
\citep{GLB08,JeeKomatsuSuyu2015,JeeEtal2016,Shajib2017}. Thus, we
provide the model deflector velocity dispersion in addition to the
\hst\ images, time delays and \kext.  An integrated line-of-sight
velocity dispersion is computed by weighting the velocity field by the
surface brightness in a square aperture with 1$''$ on a side. Typical
seeing condition is rendered by convolving the
surface-brightness-weighted line-of-sight velocity dispersion image
with a Gaussian kernel with a full width at half maximum (FWHM) of
0\farcs6.
%
%
%
%


Following current practice \citep{Shajib2017, h0licow4}, a random Gaussian noise with 5\% standard deviation is added to the model velocity dispersion to account for typical measurement errors.

\subsection{Rungs}
\label{sec:rung}

Lens modelling is usually time-consuming both in terms of human and
computer time. Thus, the size of simulated samples is limited by
practical considerations. Based on the experience of the evil team and
consultations with members of the lensing community a sample size of
50 was considered a good compromise between practicality and the need
to explore different conditions with sufficient statistics to uncover
potential biases. Thus, we construct the challenge in the following
manner. Similar to the time delay challenge we provide an entry
level zeroth rung for format checking and testing purposes. The zeroth
rung consists of two simple lenses. If the teams can successfully
recover \hc\ from the zeroth rung, they are
encouraged to participate and submit their results for 
rungs consisting of the real challenge. Each rung consists of 16
lenses. For each rung, sixteen systems are simulated including
\textit{cusp, fold, cross} and \textit{double} configurations; four
examples for each configuration are generated by using two independent codes. 

Considering the quality of the data simulated here, constraints
on \hc\ with a precision of $\sim$6\% should be possible and thus 48 
systems would deliver \hc\ to sub-percent precision which would be
sufficient to uncover biases at this level.
We set a global value of \hc\ per rung.
To ensure that the ``Good'' Teams do not infer any information for the
previous rung, we reset \hc\ at each rung.
The complexity and realism
of the systems increase with rung level, thus allowing us to separate
different aspects of the lens models and understand what needs
improvement. 
Partial submissions for a subset of the rungs will be
accepted. 

Detailed information for each rung's design including the lens
components and data provided to the ``Good'' team is given in the following subsections.

\subsubsection{Rung~0}
\label{sec:rung0}

Rung~0 is a training exercise which consists of two lens systems, one two-image (\textit{double}) and one four-image (\textit{quad}) configuration. The goal of this rung is to ensure that ``Good'' Team members understand the format of the data and that no bugs or mistakes could potentially affect the results of the challenge for a specific method.

In view of this goal, a parametric models for surface brightness model and mass profile are selected. We adopt a single S\'ersic profile to describe the surface brightness of both the lens and source galaxy, and elliptical power-law models for the lens surface mass density. Also, we randomly added an external shear to the lens potential drawn from a typical range. The AGN images are added as a point sources and the PSF is provided. The lens parameters and cosmological parameters for the simulations are released with the data for the modelling team to check.

\subsubsection{Rung~1}
\label{sec:rung1}
This rung is meant to be the easiest one of the actual challenging ladder. Thus, the mocks in
Rung~1 are generated in a similar way as in Rung~0, except that we use the images of real galaxies
to get realistic surface brightness distribution for the lensed AGN host
and the time delays are affected by external convergence (i.e., Section~\ref{kext}).
For Rung~0-1, we also provide an oversampled PSF; the pixel size is 0\farcs13/4=0\farcs0325.
This is to mimic the oversampling that
is generally achieved by combining several stars in the science image.

\subsubsection{Rung~2}
\label{sec:rung2}
Rung~2 is meant to test PSF reconstruction features of lensing codes,
in addition to the aspects tested in Rung~1. For this purpose, we only
provide a guess of the PSF but not the one actually used to generate
the data.

\subsubsection{Rung~3}
\label{sec:rung3}
Rung~3 is the highest level in this challenge, and thus the simulations are intended to be the most realistic. In addition to all the complexity we have adopted for Rung~1 and Rung~2, the observables are generated using massive early-type galaxies selected from numerical cosmological simulations. 

\section{Instructions for Participation and Evaluation Metrics}
\label{sec:participation}

Access to the simulated lens data is through the following website:
\begin{itemize}[noitemsep]
\item \url{https://tdlmc.github.io}
\end{itemize}

The ``Good'' Teams are asked to submit their point estimates of \hc\
($\tilde{H}_{0}$) and the estimated 68\% uncertainties 
($\delta$), in the format provided at the website. Multiple entries corresponding to different choices of point-estimators and credibility intervals (e.g. maximum likelihood, mean posterior) are accepted.

\subsection{Metrics}

Following \cite{TDC1,TDC2}, the ``Evil'' Team will compute
four standard metrics to measure the precision and accuracy. The
metrics will be computed for each rung and for the combined three
rungs.

The first metric is \textit{efficiency} which quantifies the
fraction of successfully modelled lenses in each rung:

\begin{eqnarray}
 \label{eq:efficiency}
f = \frac{N}{N_{total}},
\end{eqnarray} 
where  $N$ is the number of successfully modelled systems in each submission and $N_{total} = 16.$
Defining this metric means the ``Good'' team do not have to submit the result for each system, but can choose to omit the ones they cannot confidently model.
Note that the high efficiency does not necessarily map into a precise
and accurate measurement of \hc, since the removal of outliers could be an
effective way to avoid catastrophic errors.

The second metric, aiming at evaluating the \textit{goodness} of the error estimate, is the standard reduced $\chi^2$:
\begin{eqnarray}
 \label{eq:goodness}
\chi^2=\frac{1}{N} \sum_i\biggl( \frac{\tilde{H}_{0~i}-H_{0}}{\delta_i} \biggl)^2 ,
\end{eqnarray} 
where the \hc\ is the true value adopted in each rung, and ${\delta_i}$ is the uncertainty ($1-\sigma$ level) of \hc\ by each systems in the submission.

The third metric is the \textit{precision}, defined as:
\begin{eqnarray}
 \label{eq:precision}
P = \frac{1}{N} \sum_i \frac{\delta_i}{H_{0}},
\end{eqnarray} 
measuring the average relative uncertainty.

Finally, the fourth metric is the \textit{accuracy} or \textit{bias} of the estimator, which we quantify with the fractional residual:
\begin{eqnarray}
 \label{eq:accuracy}
A = \frac{1}{N} \sum_i \frac{\tilde{H}_{0~i}-H_{0}}{H_{0}}.
\end{eqnarray} 

We expect the metrics to meet the following ranges for the full challenge: 
\begin{eqnarray}
   0.6<\chi^2 <1.5,\\
   P< 6\% ,\\
   |A| < 1 \%,
\end{eqnarray}
where the $\chi^2$ range corresponds approximately to the 1 and 99
percentile of the $\chi^2$ distribution for 48 degrees of freedom
according to statistics (Rung~1-3 have 48 systems in total).  The
target for precision is based on the best results obtained so far in
the literature with data on comparable quality, while the target for
$A$ is set by our goal of sub-percent accuracy. We do not set a target
for efficiency even though of course low efficiency will be implicitly
penalized by small number statistics.  For a single rung, the $\chi^2$
range and accuracy are expected to be less stringent:

\begin{eqnarray}
   0.4<\chi^2 <2,\\
   P< 6\% ,\\
   |A| < 2 \%.
\end{eqnarray}

The metrics are analyzed individually in each rung. We expect that the performance will drop off climbing up the ladder.

\subsection{Timeline and Publication of the Results}

The challenge mock data was released on January 8th, 2018.  The deadline for blind submission of the Rung~1 is August 8th, 2018, seven month after the data release. The deadline for Rung~2 is 2019 April 8 and the one for Rung~3 is 2019 September 8. In order to allow for correction of bugs, and to test different algorithms, multiple submissions are accepted. The true input parameters will be published after the deadline of each rungs to allow teams to study in more detail their submissions, and/or for non-blind submissions.

This paper has been posted on the arXiv since January 2018 as a means to open the
challenge. After the deadline, the plan is to submit this paper to the
journal together with the second paper of this series presenting the
results and including all ``Good'' Team members as co-authors. The two
papers will be submitted concurrently so as to allow the referee to
evaluate the entire process. ``Good'' Teams are encouraged to publish
papers on their own methods using the challenge data, after the
submission of paper II. By participating in the challenge, the ``Good'' Teams agree to not publish the detailed results of their own method before the collective paper II is submitted to the journal.



\section{Summary}
\label{sec:sum}
We presented the time delay lens modelling challenge. The structure of the challenge is as follows. The ``Evil'' Team produced a set of mock lenses, meant to mimic state of the art data quality. Anyone in the community is invited to participate as a ``Good'' Team, by modelling the data and submitting a blind estimate of the Hubble Constant.  The ``Evil'' Team will compute four metrics aimed at quantifying the accuracy and precision of the estimates. The overall goal of the challenge is to assess whether current lens modelling techniques are sufficient to ultimately reach a 1\% measurement of \hc. The challenge is organized in rungs in order to help identify aspects of the lens modelling effort that may represent bottlenecks and may require additional improvements.

\section*{Acknowledgments}

We thank Vivien Bonvin, Simon Birrer, Matthew W. Auger, Xiao-Lei Meng for useful suggestions and technical supports. X.D. acknowledges support by China Postdoctoral Science Foundation Funded Project; he is also grateful for Zong-Hong Zhu's support and funding.
T.T. acknowledges support by the Packard Foundation in the form of a Packard Research Fellowship.
T.T. and C.D.F. acknowledge support by NSF through grant ``Collaborative Research: Toward a 1\% Measurement of The Hubble Constant with Gravitational Time Delays'' AST-1906976.
Xu D. acknowledges support of the Klaus Tschira foundation at HITS.
This work was supported by World Premier International Research Center Initiative (WPI Initiative), MEXT, Japan.
G.C.-F.Chen acknowledges support from the Ministry of Education in Taiwan via Government Scholarship to Study Abroad.
C.D.F. acknowledges support from the US National Science foundation grant AST-1312329.
C.D.F. and G.C.-F.C. acknowledge support for this work from the National Science Foundation under Grant Numbers AST-1312329 and AST-1907396.
S.H. acknowledges support by the DFG cluster of excellence \lq{}Origin and Structure of the Universe\rq{} (\href{http://www.universe-cluster.de}{\texttt{www.universe-cluster.de}}).
This project has received funding from the European Research Council (ERC) under the European Union's Horizon 2020 research and innovation programme (grant agreement No 787886). 
S.V. has received funding from the European Research Council (ERC) under the European Union's Horizon 2020 research and innovation programme (grant agreement No 758853).
S. Hilbert acknowledges support by the DFG cluster of excellence `Origin and Structure of the Universe'.

\section*{Data Availability}
The data underlying this article are available in the TDLMC website, at \url{https://tdlmc.github.io/}



\bibliographystyle{mnras}
\input{manuscript.bbl}








\bsp	
\label{lastpage}
\end{document}